\documentclass[aps,prb,twocolumn,showpacs,groupedaddress]{revtex4}

\usepackage{latexsym}
\usepackage[english]{babel}
\usepackage{graphicx}
\usepackage{amsfonts}
\usepackage{amsmath}
\usepackage{amssymb}
\usepackage{slashed}
\usepackage{verbatim}
\usepackage{feynmf}
\usepackage{mathrsfs}
\usepackage{epsfig}
\usepackage{relsize}
\usepackage{bbm}
\usepackage{graphics}
\usepackage{tabularx,float}
\usepackage{subfigure}
\usepackage{bm}
\usepackage{wasysym}
\usepackage{color}

\usepackage{cancel}

\newcommand{\bk}{{\bf k}}
\newcommand{\bkp}{{\bf k'}}
\newcommand{\bp}{{\bf p}}
\newcommand{\bQ}{{\bf Q}}
\newcommand{\bchi}{{\boldsymbol{\chi}}}
\newcommand{\bbeta}{{\boldsymbol{\eta}}}

\begin{document}

\date{\today}

\author{D. Makogon$^1$, R. van Gelderen$^1$,  R. Rold\'{a}n$^2$ and C. Morais Smith$^1$}
\affiliation{\centerline{$^1$Institute for Theoretical Physics, Utrecht University, Leuvenlaan 4, 3584 CE Utrecht, The Netherlands}
\centerline{$^2$Institute for Molecules and Materials, Radboud University Nijmegen, Heyendaalseweg 135, 6525 AJ Nijmegen, The Netherlands}}

\title{Spin-density-wave instability in graphene doped near the van Hove singularity}

\begin{abstract}
We study the instability of the metallic state towards the formation of a new ground state in graphene doped near the van Hove singularity. The system is described by the Hubbard model and a field theoretical approach is used to calculate the charge and spin susceptibility. We find that for repulsive interactions, within the random phase approximation, there is a competition between ferromagnetism and spin-density wave (SDW). It turns out that a SDW with a triangular geometry is more favorable when the Hubbard parameter is above the critical value $U_c (T)$, which depends on the temperature $T$, even if there are small variations in the doping. 
Our results can be verified by ARPES or neutron scattering experiments in highly doped graphene.
\end{abstract}

\pacs{72.20.-r, 73.22.-f, 75.30.Fv, 75.70.Ak}

\maketitle

\section{INTRODUCTION}

Graphene, a newly realized two-dimensional crystal of carbon atoms ordered on a honeycomb lattice,\cite{NF04} is being extensively studied due to its unusual electronic and structural properties. Undoped graphene is a zero-gap semiconductor with a linear low-energy dispersion relation and a vanishing density of states at the Fermi level.\cite{CaNe09a} Although the Coulomb interaction is unscreened in this regime, the system behaves mostly as a non-interacting electron liquid, where the minor effects of interactions are encoded in a renormalization of the Fermi velocity and the quasiparticle lifetime.\cite{GGV94} The opposite case is that of a divergent density of states, the so called Van Hove (VH) singularity, which is associated to a saddle point in the band dispersion.\cite{VH53} If the Fermi level lies near such a singularity, screening is perfect for wave-vectors connecting VH singularities and correlation effects may be enhanced. Recently, several experimental groups have succeeded on making this regime accessible for graphene.\cite{LA09,MR10,CS10} The techniques used for this aim are different: in Ref.\ \onlinecite{LA09} they have used twisted graphene, obtained from a rotation between stacked graphene layers, which allows one to tune the position of the VH singularity. Other methods involve chemical doping of graphene\cite{MR10} or intercalation of gold clusters between the graphene layers.\cite{CS10} This opens up the fascinating possibility to study correlated electronic phases in this material, such as superconductivity, charge-density wave (CDW) or spin-density wave (SDW).

The proximity of the Fermi level to a VH singularity in most of the cuprate superconductors has  triggered large efforts to understand the role of a peaked density of states (DOS) on the electronic properties of an electron liquid.\cite{M97} Superconductivity, itinerant ferromagnetism (FM),  CDW,  and SDW are examples of competing instabilities associated with the VH scenario. In graphene, the existence of some of these instabilities has already been evidenced by recent experiments, such as CDW\cite{LA09,CS10} or superconducting pairing due to electron-electron interactions,\cite{MR10} following the Kohn-Luttinger mechanism.\cite{KL65}
CDW and SDW phases usually occur in systems with Fermi surface nesting, i.e. when the Fermi surface can
be mapped onto itself by a (nonzero) $\bk$-vector (nesting vector).
At low doping, graphene shows some nesting, since the Fermi surface is composed of two circles around the $K$ and $K'$ points. However, since the DOS is very low, these peaks are small and a very high (Hubbard) interaction is required to enter the DW regime.\cite{Nuno} The situation changes if we strongly dope graphene around the value $\mu \approx |t|$, where $\mu$ is the chemical potential and $t\approx 2.8$eV is the hopping parameter. The Fermi surface then acquires a hexagonal
 shape and nesting occurs in three directions (see Fig.\ \ref{fig1}). Furthermore, the VH singularities lie at the Fermi surface for this doping. At these points, the DOS diverges and we expect some non-trivial peaks in the susceptibility.

In this paper, we investigate the Hubbard model in a honeycomb lattice using a field theoretical approach. By performing a Hubbard-Stratonovich transformation, we determine the effective action in terms of an 8-component order parameter, which accounts for charge- and spin-degrees of freedom in the A and B graphene sublattices. We find that the charge susceptibility never diverges for repulsive interactions, thus excluding the possibility of a CDW formation, whereas the spin susceptibility exhibits several peaks. The peaks at the nesting wavevectors $\bk$ are stronger than the one at $\bk = 0$ (see Fig.\ \ref{fig2}), thus signaling that SDW is the leading instability, that wins against FM for repulsive interactions.

In the following, we introduce the model and outline the main steps of the calculations in Sec. II. Then, we present the temperature vs interaction vs doping phase diagram and show that the metallic state becomes unstable towards a SDW phase for realistic values of the interaction parameter $U$ in Sec. III.
Finally, we discuss possible experimental techniques which could  be used to observe our results and draw our conclusions in Sec. IV.

\begin{figure}[t]
\includegraphics[width=.35\textwidth, bb= 0 0 1113 955]{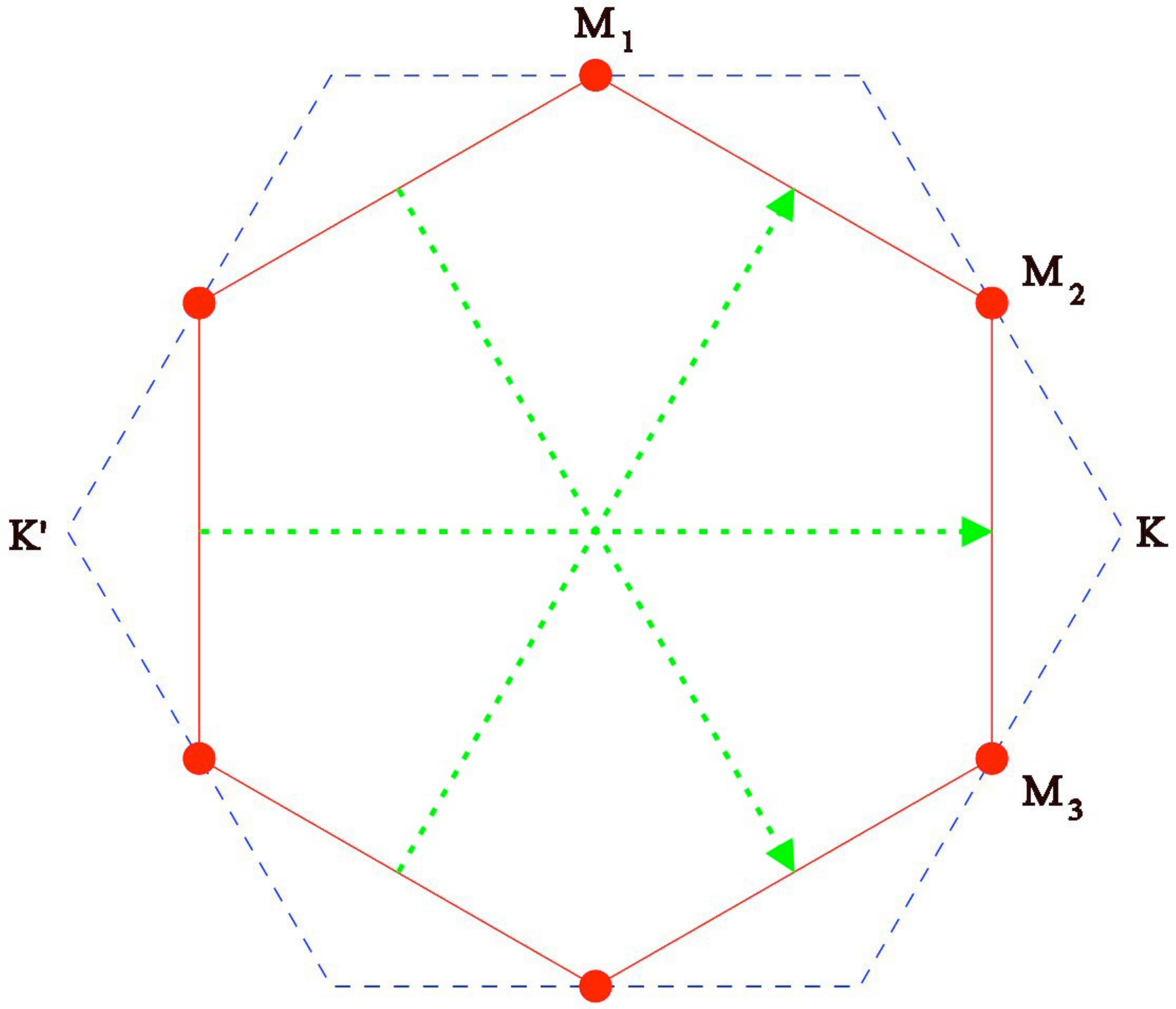}
\label{fig1}
\caption{(Color online) (a): Blue dashed line is the Brillouin zone of monolayer graphene. Red line is the Fermi surface for $\mu=t$. Red dots are the van Hove singularities and the green arrows are the nesting vectors.
}
\label{fig1}
\end{figure}

\section{THE MODEL}

Due to the strong screening of interactions by the electron gas at the VH filling, we consider only the local Coulomb repulsion (the usual Hubbard U term), which is a good approximation around this doping level.\cite{GGV97,RLG08,ST10}
This term leads to a deformation of the band dispersion toward the saddle point,\cite{RLG08} an effect that has been observed experimentally by ARPES.\cite{MR10} Notice that longer-range interaction terms, such as the nearest-neighbor repulsion, would lead to a richer phase diagram, with the possibility of a Pomeranchuk instability.\cite{VV08,LCG09}
This scenario is in sharp contrast with low doped graphene, for which the DOS vanishes and the Coulomb interaction will be only slightly screened and therefore will be long ranged.
Hence, we use a tight-binding model with a Hubbard interaction $U$,
\begin{equation}
H=-t \sum_{<i,j>,s} (a^\dagger_{i,s} b_{j,s} +H.c.) + U \sum_{j \in A, B} c^\dagger_{j, \uparrow} c^\dagger_{j,\downarrow} c_{j, \downarrow} c_{j, \uparrow},
\label{eq1H}
\end{equation}
where
the operator $c_j$ can be either $a_j$ or $b_j$, depending whether $j$ is a label of the $A$ or $B$ sublattice, respectively. Defining $c^\dagger_j=(c_{j,\uparrow}^\dagger,c^\dagger_{j,\downarrow})$, the Hubbard term can be rewritten using the relation,
$$ c^\dagger_{j, \uparrow} c^\dagger_{j,\downarrow} c_{j, \downarrow} c_{j, \uparrow}=\frac{1}{8} n_j^2-\frac{1}{2} \mathbf{S}_j \cdot \mathbf{S}_j,$$
where $n_j=c^\dagger_j c_j$ is the on site number operator and $\mathbf{S}_j=(1/2) \, c^\dagger_j \, \boldsymbol{\sigma} \, c_j$, describes the spin on the lattice site $j$. Note that the inner product $\mathbf{S}_j \cdot \mathbf{A}$, where $\bf A$ is some vectorial field, is defined by $\mathbf{S}_j \cdot \mathbf{A}=(1/2)\, c^\dagger_j \, (\boldsymbol{\sigma} \cdot \mathbf{A}) \, c_j$, and that $\boldsymbol{\sigma}$ is the vector of Pauli matrices.
The grand-canonical partition function describing the system reads $$Z=\int \textrm{d}[c^\dagger] \textrm{d} [c] e^{-S[c^\dagger,c]/\hbar},$$ where the action is given by $$S=\int_0^{\hbar \beta} \textrm{d} \tau \left[ \sum_j c^\dagger_j \left( \hbar \frac{\partial}{\partial \tau}-\mu \right) c_j+H \right]. $$
In this expression, $\tau$ is the imaginary time variable, $\beta=1/k_BT$
and $H$ is defined in Eq.~(\ref{eq1H}).
We will investigate the possible appearance of CDW and SDW instabilities by using a path integral formalism recently developed by some of the authors.\cite{Dima10a}
In the following, we briefly outline the procedure.

We start by performing a Hubbard-Stratonovich transformation that eliminates the quartic term in the action, but introduces eight auxiliary bosonic fields $\rho^a$, $\rho^b$, $\mathbf{M}^a$, and $\mathbf{M}^b$ related, respectively, to the electronic and the magnetic degrees of freedom of the fermionic fields of each sublattice. In Fourier space, the action then reads
\begin{align}
\nonumber S &= -\hbar \sum_{\bk,n,\bkp,n'} \left( a^\dagger_{\bk,n} b^\dagger_{\bk,n} \right)\cdot \mathbf{G}^{-1}_{\bk,n;\bkp,n'} \cdot \left( \begin{array}{c} a_{\bkp,n'} \\ b_{\bkp,n'} \end{array} \right) + S_2, \\
\nonumber S_2 &= \frac{1}{2U} \sum_{\bk,n,\alpha} \left[ \mathbf{M}^\alpha_{\bk,n} \cdot \mathbf{M}^\alpha_{\bk,n}-(\rho^\alpha_{\bk,n})^2 \right],
\end{align}
where $\alpha=a,b$. The inverse Green's function is defined by $ \mathbf{G}^{-1}_{\bk,n;\bkp,n'} =\mathbf{G}^{-1}_{0 \, \bk,n;\bkp,n'}-\boldsymbol{\Sigma}_{\bk,n;\bkp,n'}$, where the bare Green's function reads
\begin{align}
\nonumber -\hbar \mathbf{G}^{-1}_{0 \, \bk,n;\bkp,n'} &= \left[ \begin{array}{cc} -(\mu+ i\hbar \omega_n) \mathbf{I} & -t \gamma_{\bk} \mathbf{I} \\ -t \gamma_{\bk}^* \mathbf{I} & -(\mu+i \hbar \omega_n) \mathbf{I} \end{array} \right] \delta_{\bk,\bkp} \delta_{n,n'},
\end{align}
with $\mathbf{I}$ a $2 \times 2$ identity matrix, $\omega_n = \pi(2n+1)/\hbar\beta$ the fermionic Matsubara frequency, and $\gamma_{\bk} =e^{i a_0 k_y}+e^{-i a_0 k_y/2}\cos(\sqrt{3}a_0 k_x/2)$. In what follows we set the lattice constant $a_0=1$. The self energy is given by
\begin{align}
\nonumber \hbar \boldsymbol{\Sigma}_{\bk,n;\bkp,n'} &= \frac{-1}{2 \sqrt{N \hbar \beta}} \left[ \begin{array}{cc} \boldsymbol{\sigma} \cdot \mathbf{M}^a-\rho^a \mathbf{I} & 0 \\ 0 & \boldsymbol{\sigma} \cdot \mathbf{M}^b-\rho^b \mathbf{I} \end{array} \right],
\end{align}
where $N$ is the number of sites of each sublattice and both, ${\bf M}^\alpha$ and $\rho^\alpha$ carry a subindex $(\bk-\bkp,n-n')$. By introducing the eight component vector
$$\mathbf{M}_{\bk,n}=\left( \rho_{\bk,n}^a, \\ \mathbf{M}_{\bk,n}^a, \\ \rho^b_{\bk,n}, \\ \mathbf{M}^b_{\bk,n} \right)^{\rm T}$$
and integrating out the fermionic fields, we obtain the effective action
$$S_\textrm{eff}=\frac{1}{2U} \sum_{\bk,n} \mathbf{M}_{\bk,n} \cdot \bbeta \cdot \mathbf{M}_{-\bk,-n}-\hbar \textrm{Tr}[ \ln(-\mathbf{G}^{-1})],$$
where $\bbeta=\textrm{diag}(-1,1,1,1,-1,1,1,1)$ and the partition function becomes $Z=\int \textrm{d}[\mathbf{M}] \exp(-S_\textrm{eff}/\hbar)$. Finding the susceptibility from here is a two steps process. First, we introduce external source fields $\mathbf{J}$ and then we expand the fields in the action around their mean field value $\mathbf{M}_{\bk,n}= \langle \mathbf{M}_{\bk,n} \rangle_\mathbf{J}+\delta \mathbf{M}_{\bk,n}$. This results in a self-consistent equation for the mean field values of the fields. Using this equation, we may determine how the bosonic fields react on a distortion of the source fields.


\begin{figure}[t]
\includegraphics[width=.45\textwidth]{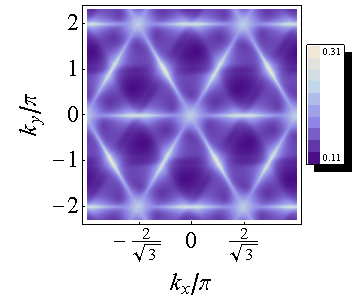}
\caption{(Color online) Density plot of the largest eigenvalue of $\bbeta \boldsymbol{\chi}$ for $T = 0.01 t$. Brighter regions correspond to higher peaks in the spin susceptibility.
The peaks at the three nesting wavevectors (see Fig.\ \ref{fig1}) are higher than the one at ${\bf k} = 0$, indicating that the SDW instability wins over the FM one. }
\label{fig2}
\end{figure}

The mean field values are those at which the action has an extremum. Thus, the linear terms in $\delta \mathbf{M}_{\bk,n}$ are imposed to vanish, yielding the self-consistent equation
$$ \langle M^r_{\bk,n} \rangle_\mathbf{J}=\frac{U}{2 \sqrt{N \hbar \beta}} \sum_{\bp,m} \textrm{Tr}[\mathbf{P}^r \mathbf{G}_{\mathbf{J} \, \, \bp,m;\bp-\bk,m-n}],$$
where  $\mathbf{P}=[ \textrm{diag}(\mathbf{I},0),\textrm{diag}(\sigma^x,0),\textrm{diag}(\sigma^y,0),\textrm{diag}(\sigma^z, 0),  \\ \textrm{diag}(0,\mathbf{I}),
\textrm{diag}(0,\sigma^x),\textrm{diag}(0,\sigma^y),\textrm{diag}(0,\sigma^z)]^{\rm T}$
and the trace is taken over the matrices only, because the sum takes care of the trace in $\bk$- and frequency-space.
After differentiating this equation with respect to the source $\mathbf{J}$, one finds the susceptibility through the relation
\begin{align} \label{dmchidj} \textrm{d}\langle\mathbf{M}\rangle_\mathbf{J}=\hbar U \bchi^\textrm{RPA}_\mathbf{J} \textrm{d} \mathbf{J},
 \end{align} where $$\bchi^\textrm{RPA}_\mathbf{J}=\left( \mathbf{I}-U \bchi_\mathbf{J} \bbeta \right)^{-1} \bchi_\mathbf{J}.$$
Moreover, we can also determine the free energy by performing
a Legendre transformation $\beta F[\langle\mathbf{M}\rangle_{J}]=
U^{-1}\langle\mathbf{M}\rangle^{\dag}_{J} \cdot
\mathbf{J}-\ln(Z[\mathbf{J}])$ on the partition function which, up
to quadratic order in the deviation
$\Delta\langle\mathbf{M}\rangle_{J}\equiv\langle\mathbf{M}\rangle_{J}-\langle\mathbf{M}\rangle_{0}$
and without an additive constant, is
\begin{equation}\label{quadraticfreeenergy}
\beta F[\langle\mathbf{M}\rangle_{J}]=\frac{1}{2\hbar U^2}\left(
\Delta\langle\mathbf{M}\rangle_{J}\right)^{\dag} \cdot
(\mathbf{\chi}_{0}^{\rm RPA})^{-1}
\cdot\Delta\langle\mathbf{M}\rangle_{J}.
\end{equation}
The susceptibility $\mathbf{\chi}_{0}^{\rm RPA}$ is evaluated in
the absence of the source field, $\mathbf{J}=0$.
From here on, we assume that the system is in a homogeneous state, for which the mean field values of the boson fields are given by $\langle M^r_{\bk,n} \rangle_0=\langle M^r_{0,0} \rangle \delta_{\bk,0} \delta_{n,0}.$ As a consequence, the self-energy, the Green's function and the susceptibility matrix $\bchi_0$ are all diagonal in momentum and frequency space.
Due to a nonzero self-energy, the Hamiltonian gets renormalized, such that $$\hbar \mathbf{G}^{-1}=(\mu+i \hbar \omega_n) \mathbf{I}-H_0-\boldsymbol{\Sigma} \equiv (\mu+i \hbar \omega_n) \mathbf{I}-H_\textrm{ren},$$
where $H_0$ is given by the first term in Eq.\ (1).
Assuming a (renormalized) Hamiltonian, which may be diagonalized by using the unitary operators ${\cal U}_{\bk}$, such that
$ {\cal U}^\dagger_{\bk} H_{\bk} {\cal U}_{\bk}=\sum_\alpha I^{(\alpha)} \epsilon_{\bk}^{(\alpha)},$ where $I^{(\alpha)}= \textrm{diag}(\delta_{\alpha,1},\delta_{\alpha,2},\delta_{\alpha,3},\delta_{\alpha,4})$ and
$\tilde{\epsilon}^{(\alpha)}_{\bk}=\epsilon^{(\alpha)}_{\bk}-\mu$, the susceptibility can be written as \cite{Dima10a}
\begin{align}
\nonumber \chi_{\bk}^{r,r'}(i \hbar \Omega_n)&= \frac{1}{N} \sum_{\bp,\alpha,\beta} \frac{n_F\left(\tilde{\epsilon}^{(\alpha)}_{\bp+\bk}\right)-n_F\left(\tilde{\epsilon}^{(\beta)}_{\bp}\right)}{\tilde{\epsilon}^{(\alpha)}_{\bp+\bk}-\tilde{\epsilon}^{(\beta)}_{\bp}-i \hbar \Omega_n} T^{r,r';\alpha, \beta}_{\bp+\bk,\bp}, \\
\nonumber T^{r,r';\alpha, \beta}_{\bp+\bk,\bp}&= -\frac{1}{4} \textrm{Tr}[P^r {\cal U}_{\bp+\bk} I^{(\alpha)} {\cal U}^\dagger_{\bp+\bk} P^{r'} {\cal U}_\bp I^{(\beta)} {\cal U}^\dagger_{\bp}],
\end{align}
where $\Omega_n$ are bosonic Matsubara frequencies and $n_F (x)=1/(e^{\beta x}+1)$ is the Fermi distribution function. Notice that the overlap between the electron and hole wave-functions is taken into account through the term $T^{r,r';\alpha, \beta}_{\bp+\bk,\bp}$.
\begin{figure}
\includegraphics[width=.35\textwidth]{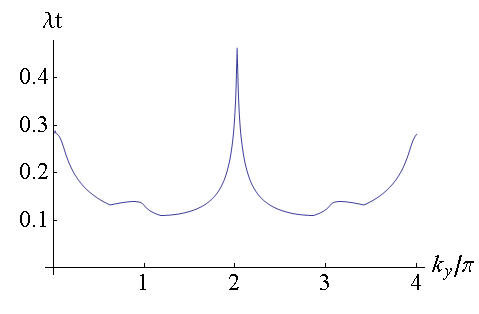}
\caption{(Color online) Plot of the largest eigenvalue of $\bbeta \bchi$ for $T = 0.0025 t$ for $k_x = 0$. Notice that the peak corresponding to SDW at $k_y = 2 \pi$ is larger than the FM peaks at $k_y = 0$ and $k_y = 4 \pi$. } 
\label{fig3}
\end{figure}
\begin{table}[htb]{
\begin{tabular}{|l|l|l|l|}
 \hline $T/t (10^{-3})$ & T (K) &  $U^\textrm{FM}_c$ &  $U^\textrm{SDW}_c$ \\
 \hline
 \hline 0.025 &0.87& 1.0570 & 0.8444 \\
 \hline 0.05 &1.74& 1.6710 & 0.9954 \\
 \hline 0.075 &2.6& 2.0410 & 1.0776 \\
 \hline 0.15 &5.2& 2.5018 & 1.2184 \\
 \hline 0.25 &8.7& 2.6828 & 1.3322 \\
 \hline 0.5 &17& 2.8896 & 1.5118 \\
 \hline 1 &35& 3.1272 & 1.7288 \\
 \hline 2.5 &87& 3.5082 & 2.0920 \\
 \hline 5 &175& 3.8652 & 2.4444 \\
 \hline 10 &350& 4.3030 & 2.8878 \\
 \hline
\end{tabular}
}
\caption{Numerical values of the critical coupling $U_c$ (in units of $t$) for a FM and a SDW phases calculated on a 10000$\times$10000 mesh. }
\label{tab1}
\end{table}
\begin{figure*}[t]
\subfigure[]{
\includegraphics[width=.3\textwidth]{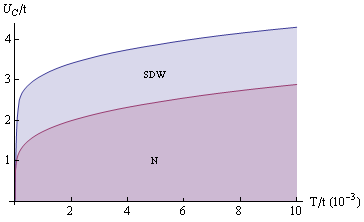}
\label{fig4a}
}
\subfigure[]{
\includegraphics[width=.3\textwidth]{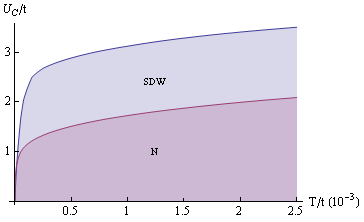}
\label{fig4b}
}
\subfigure[]{
\includegraphics[width=.3\textwidth]{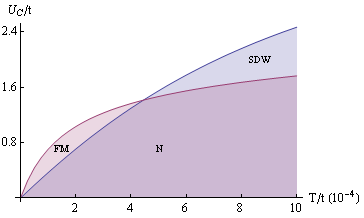}
\label{fig4c}
}
\caption{(Color online) (a) Plot of the critical coupling required to enter the SDW regime (red line) and the FM regime (blue line) as a function of temperature, starting from the metallic phase at small U. We observe that the SDW phase wins for all temperatures. (b) Zoom-in of figure (a) for low temperatures. Both figures have been determined using a 10000$\times$10000 mesh, for which convergence was eventually reached for $T > 0.5 \times 10^{-3}$. (c) Same as (a), but calculated in a 700$\times$700 mesh. When not enough points are taken into account, the two lines cross, thus leading to an erroneous conclusion that a FM phase would be more favorable at low temperatures. }
\label{fig4}
\end{figure*}

\section{NUMERICAL RESULTS}

Next, we investigate the possible instabilities by considering the static susceptibility $\bchi_{\bk}(0)$.
To evaluate the latter, one needs to perform a sum over all momenta in the Brillouin zone. We proceed by dividing the Brillouin zone in an $N \times N$ mesh, which implies that the sum in the expression for the susceptibility has $N^2$ terms.
The instability condition for repulsive interactions requires the interaction strength $U$ to exceed the critical value $U_c$ defined by $0=\det (\bchi_{J=0}^{-1}-U_c \bbeta)=\det(\bbeta \bchi_{J=0}^{-1}-U_c).$ This relation links the critical interaction strength to the largest eigenvalue $\lambda_{\bk}$ of the matrix $\bbeta \bchi_{\bk}(0)$ by $$U_c^{-1}=\max_{\bk} \lambda_{\bk} \equiv \lambda_{\bQ}.$$ In Fig.~\ref{fig2} a density plot of the largest eigenvalue of the matrix $\bbeta \bchi_{\bk}(0)$ is shown for a temperature of $T=0.01 t \approx 350$ K. One can distinguish four inequivalent peaks (bright regions in Fig.~\ref{fig2}). The one at ${\bk}=0$ corresponds to a FM instability, while the other three, at nonzero ${\bk}$, correspond to the three independent nesting vectors (see Fig.~\ref{fig1}) and hence give rise to a SDW. Although not expected, for all temperatures the SDW peaks turn out to be higher than the peak corresponding to  the FM phase (see Fig.~\ref{fig3}).  However, this result depends heavily on the size of the mesh: for low temperatures, the sum converges slowly, and a wrong result can be inferred if $N$ is not large enough. Indeed, by considering a $700 \times 700$ mesh, we find that a FM phase would set in at low temperatures [see Fig.~\ref{fig4c}], whereas the results for a finer mesh ($10000 \times 10000$) indicate that the true ground state is a SDW [see
Fig.~\ref{fig4a} and Fig.~\ref{fig4b}].
\begin{figure}[b]
\includegraphics[width=.45\textwidth, bb=0 0 360 283]{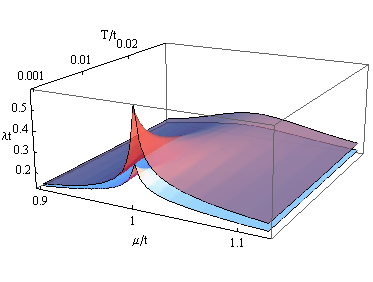}
\caption{Largest eigenvalue vs temperature and vs chemical potential. The critical $U$ scales as $1/\lambda_{\bQ}$ and will therefore increase if the doping level is tuned away from $\mu=|t|$.}
\label{fig5}
\end{figure}

For zero temperature and a doping value exactly at the VH singularity, the critical coupling for both FM and SDW is zero. As explained below, our formalism is not suited to determine which state is more favorable. For finite temperatures, the SDW phase transition has a lower critical $U$ than the FM one. Therefore, between the red and blue lines in Fig.~\ref{fig4} we find a SDW phase. Above the blue line, the system allows for a FM phase transition. The starting point of our formalism is a homogeneous ground state, after which we can determine the critical $U$ that is needed to enter a more ordered phase, like the FM, SDW or CDW phase. To reach the regime above the blue line in Fig.~\ref{fig4}, one has to start from a SDW ground state and subsequently determine if a phase transition to the FM phase will occur. Since the SDW ground state is inhomogeneous, our formalism is thus unable to determine the leading instability when $U$ is in this regime.

If the system is tuned away from optimal doping, the VH singularities will no longer lie on the Fermi surface and nesting is reduced in general. This will result in a lowering of the peaks in the maximal eigenvalue of the susceptibility and hence the critical couplings will increase. This behavior is shown in Fig.~\ref{fig5}. The SDW peak is always higher than the FM one, but the height decreases quite rapidly as function of doping. The interaction strength in graphene cannot be tuned externally. Although at low doping the onsite Hubbard interaction is estimated to be around $U/t=3.5$,\cite{CaNe09a,WeBl11} values of $U$ for highly doped graphene are, to the best of our knowledge, not yet known, but should not be different from the low-doping values. We therefore expect that a SDW phase should be observed experimentally, in the regime of $T$ and $U$ parameters shown in Fig.~\ref{fig5} (see also Table \ref{tab1}).

\section{DISCUSSIONS AND CONCLUSIONS}
Recently, a CDW has been experimentally observed in twisted graphene bilayers doped at the VH singularity, but it has disappeared when the hopping parameter between the two layers has been put to zero.\cite{LA09} A CDW phase seems also to be present for graphene grown on top of a superlattice of gold intercalated clusters,\cite{CS10} and a
superconducting phase has been conjectured to occur for graphene heavily doped with Ca and K on both sides.\cite{MR10} However, it is unclear which is the appropriate theoretical model to describe graphene under these circumstances, and for the conventional Hubbard model in a single layer, a CDW can only occur for attractive on-site interactions.

In conclusion, we have evaluated possible SDW and CDW instabilities from a metallic phase for the honeycomb lattice of a single layer of graphene doped at the VH filling.
We found that charge and spin degrees of freedom are decoupled, and that CDWs occur for attractive, whereas SDWs occur for repulsive interactions. A peak in the spin-susceptibility at zero wave-vector has also been found, although precise calculations indicate that it is always sub-dominant with respect to the ones appearing at the nesting wavevectors, thus determining that the SDW instability must win over FM in the neighborhood of the metallic phase. This finding contradicts a previous result in the literature,\cite{Nuno} and
could only be reached after a very careful and time consuming computation in a very large mesh.
An interesting point is that the SDW phase is expected to occur at rather high temperatures. This is not very surprising, given that phenomena such as CDW or the integer quantum Hall effect, that usually occur at very low temperatures, have already been detected at high temperatures in graphene ($T = 77$K for the first and room temperature for the latter).\cite{CS10,IQHE}
We hope that our results may stimulate further experiments to unveil the existence of a SDW phase in heavily doped graphene.

We acknowledge financial support from the Netherlands Organization for Scientific Research (NWO). R. Rold\'{a}n thanks the EU-India FP-7 collaboration under MONAMI. We are indebted to Bel\'{e}n Valenzuela for a careful reading of this manuscript and for pointing out some ambiguities in the first draft of this paper.



\end{document}